\documentclass[aps,prl,preprint,showpacs]{revtex4-1}
\usepackage{amsfonts}
\usepackage{amsmath}
\usepackage{amssymb}
\usepackage{graphicx}%
\begin{document}


\title{First-principles study of intermediate-spin ferrous iron in the Earth's lower mantle}

\author{Han Hsu$^{1}$ and Renata M. Wentzcovitch$^{2,3}$}

\affiliation{$^{1}$Department of Physics, National Central
University, Jhongli City, Taoyuan 32001, Taiwan \\$^2$Department of
Chemical Engineering \& Materials Science, University of Minnesota,
Minneapolis, Minnesota, USA \\$^3$Minnesota Supercomputing
Institute, University of Minnesota, Minneapolis, Minnesota, USA}

\date{\today}

\bigskip

\begin{abstract}

Spin crossover of iron is of central importance in solid Earth
geophysics. It impacts all physical properties of minerals that
altogether constitute $\sim 95$ vol\% of the Earth's lower mantle:
ferropericlase [(Mg,Fe)O] and Fe-bearing magnesium silicate
(MgSiO$_3$) perovskite. Despite great strides made in the past
decade, the existence of intermediate-spin (IS) state in ferrous
iron (Fe$^{2+}$) (with total electron spin $S=1$) and its possible
role in the pressure-induced spin crossover in these lower-mantle
minerals still remain controversial. Using density functional theory
$+$ self-consistent Hubbard $U$ (DFT$+U_{sc}$) calculations, we
investigate all possible types of IS states of Fe$^{2+}$ in (Mg,Fe)O
and (Mg,Fe)SiO$_3$ perovskite. Among the possible IS states in these
minerals, the most probable IS state has an electronic configuration
that significantly reduces the electron overlap and the iron nuclear
quadrupole splitting (QS). These most probable IS states, however,
are still energetically disfavored, and their QSs are inconsistent
with M\"{o}ssbauer spectra. We therefore conclude that IS Fe$^{2+}$
is highly unlikely in the Earth's lower mantle.

\end{abstract}


\maketitle

\newpage

\section{Introduction}

Spin crossover, a phenomenon of interdisciplinary interest, can
occur in various length scales, including molecules (coordination complexes or coordination compounds), epitaxial thin films, and bulk solids. Transition-metal
ions with 4--7 $d$ electrons ($d^4$--$d^7$ ions) contained in these
systems can undergo a change of total electron spin ($S$) induced by
extraneous factors, such as temperature, pressure, strain, chemical
doping, or electromagnetic fields. Among the known spin-crossover
systems, the Earth's lower mantle is the largest. Located 660--2890
km deep, this region of the Earth interior has a wide pressure ($P$)
and temperature ($T$) range, spanning over 23--135 GPa and
1900--4000 K, respectively. The lower mantle is dominated by
iron-bearing minerals: $\sim20$ vol\% of ferropericlase (Fp)
[(Mg,Fe)O], $\sim75$ vol\% of Fe-bearing magnesium silicate
(MgSiO$_3$) perovskite (Fe-Pv), and a relatively thin layer of
Fe-bearing MgSiO$_3$ post-perovskite (Fe-Ppv) located in its bottom
(D" layer). Ever since the observation of spin crossover in Fp and
Fe-Pv \cite{MgFeO by Badro, XES-by-Badro}, the work on these
minerals has risen to a new high, especially for Fp, due to its
simple rock-salt structure. It is believed that Fe$^{2+}$ in Fp
undergoes a crossover from the high-spin (HS) state ($S=2$) to the
low-spin (LS) state ($S=0$) between 40--70 GPa. This spin crossover
directly affects the structural, elastic, thermodynamic, optical,
and conducting properties of Fp \cite{MgFeO_by_Antonangeli, MgFeO by
Crowhurst, MgFeO by Goncharov, MgFeO by Kantor 2006, MgFeO by Kantor
2009, MgFeO by Lin 2005, MgFeO by Lin 2007, MgFeO conduct by Lin,
MgFeO_by_Marquardt, MgFeO by Speziale 2005, MgFeO by Tsuchiya,
Wentzcovitch PNAS, MgFeO by Wu, MgFeO_Cij_PRL2013}; it also affects
iron diffusion and thus perhaps viscosity and iron partitioning in
the Earth interior \cite{MgFeO_diffusion_Ammann,
MgFeO_diffusion_Saha 2011, MgFeO_diffusion_Saha 2013}. Based on
these findings, geophysical effects of spin crossover have been
anticipated.

\bigskip

In contrast, spin crossover in Fe-Pv, the \textit{major}
lower-mantle mineral phase, has been highly controversial
\cite{Bengtson 2008, QS-by-Bengtson, QS-Pv-Catalli, Hofmeister 2006,
QS MgFeSiO3 Hsu, Ferric, QS-by-Jackson, XES-by-Li, QS-by-Li,
QS-by-McCammon, Pv_conduct_Potapkin, Stackhouse 2007, Koichiro, Site
degeneracy Umemoto, Zhang 2006}, due to the complex nature of this
mineral. In addition to Fe$^{2+}$ that substitutes Mg in the
dodecahedral ($A$) site forming (Mg,Fe)SiO$_3$ Pv, there is also
ferric iron (Fe$^{3+}$) substituting both Mg and Si [residing the
octahedral ($B$) site], forming (Mg,Fe)(Si,Fe)O$_3$ Pv. With the
recent findings made by first-principle calculations \cite{QS
MgFeSiO3 Hsu, Ferric}, a consensus has gradually been reached: only
Fe$^{3+}$ residing in the $B$ site undergoes a crossover from HS
($S=5/2$) to LS ($S=1/2$) state; iron in the $A$ site remains in HS
state, regardless of its oxidation state. The geophysical
consequences of spin crossover are still unclear, but its possible
effects on mineral properties have been reviewed or summarized in
literatures \cite{Fp Pv by Lin, Fp Pv by Lin 2013, Fp Pv by Hsu,
Fe-Al Ppv, Wentzcovitch EJM, Ppv QS by Yu}. More recently, it was
found that Fe-Pv dissociates into Fe-free Pv and a hexagonal
iron-rich silicate at conditions existing approximately at 2,000 km
depth and beyond \cite{Zhang-Mao-2014}. The crystal structure and stability field of this hexagonal phase, however, have not been characterized yet. Therefore, it is important to properly characterize the state of iron at lower mantle conditions, so the dissociation phase boundary in Fe-Pv can be better clarified.

\bigskip

While the spin crossovers in Fp and Fe-Pv are nearly understood, one
issue still remains unresolved. As a $d^6$ ion, an intermediate-spin
(IS) state with $S=1$ is possible for Fe$^{2+}$. The existence of IS
Fe$^{2+}$ in Fp and (Mg,Fe)SiO$_3$ Pv, however, has not been
\textit{fully} confirmed nor excluded. For Fp, X-ray emission
spectroscopy (XES) spectra show the total electron spin moment
decreasing with pressure, as indicated by the decreasing satellite
peak ($K\beta'$) intensity \cite{MgFeO by Badro, MgFeO by Lin 2005,
MgFeO by Lin 2007}. However, both the currently perceived HS-LS
crossover or a more complicated HS-IS-LS crossover can lead to
decreasing $K\beta'$. Also, while the change of iron nuclear
quadrupole splitting (QS) observed in M\"{o}ssbauer spectra
\cite{MgFeO by Kantor 2006, MgFeO by Kantor 2009, MgFeO by Speziale
2005} indicates a change of $d$-electron configuration, it is
insufficient to exclude or confirm an IS state. Recently, the
existence of IS Fe$^{2+}$ in Fp was investigated, but its possible
role in spin crossover was not addressed \cite{MgFeO_IS_Larico}. As
for (Mg,Fe)SiO$_3$ Pv, IS Fe$^{2+}$ has been highly debated. An
observed crossover from a lower QS ($\sim 2.4$ mm/s) to a higher QS
($\geq 3.5$ mm/s) was suggested to be indicative of an HS-IS
crossover, as the high QS was suggested to be a signature of IS
Fe$^{2+}$ \cite{QS-by-McCammon}. Previous first-principles
calculations, however, showed that two distinct types of HS states
with distinct QSs and one IS state are possible; it is the crossover
between two HS states leading to the drastic change of QS, from 2.4
to 3.5 mm/s \cite{QS MgFeSiO3 Hsu, Ferric}. The one IS state, on the
other hand, is energetically unfavorable; its QS obtained by
calculation ($\leq 1.6$ mm/s) was not observed in experiments either
\cite{QS MgFeSiO3 Hsu, Ferric}. So far, IS Fe$^{2+}$ in Fe-Pv is
still puzzling. One reason is the lack of a thorough knowledge for
the IS Fe$^{2+}$ reported in Ref.~\onlinecite{QS MgFeSiO3 Hsu}; the
other is the possibility of multiple types of IS Fe$^{2+}$. After
all, Fe-Pv is known to have two distinct types of HS Fe$^{2+}$; it
may have multiple types of IS Fe$^{2+}$ as well. Recently, an
anomalous conductivity change in Fe-Pv with increasing pressure was
observed, and it was attributed to a possible HS-IS crossover of
Fe$^{2+}$ \cite{Pv_conduct_Potapkin}. Given that the mechanism of
spin crossover is usually deduced from anomalous change in mineral
properties \textit{indirectly} related to iron spin state, a
comprehensive theoretical study for IS Fe$^{2+}$ in Fp and Pv is
highly desirable to clear this long-standing debate.

\bigskip

In addition to geophysics, IS state in Fe$^{2+}$ and other $d^6$
ions are of broad interest. Confirmation of IS Fe$^{2+}$ in minerals
can significantly change our current knowledge of the iron spin
distribution (spin map) in the Earth, and an accurate spin map can
be used to test theories beyond the standard model in particle
physics \cite{Hunter 2013}. A possible connection between IS
Fe$^{2+}$ and superconductivity in iron-based superconductors has
been discussed but is still unclear \cite{FeSC_Gnezdilov_2011,
FeSC_Kruger_2009, FeSC_Simonelli_2011, FeSC_Yu_2011}. IS Fe$^{2+}$
would add versatility to the range of possibility of molecular devices based on coordination complexes/compunds \cite{TopCurrChem}, but its existence is still controversial
(see e.g. Ref.~\onlinecite{Tracking_spin_Zhang_2014}) and conditions
for its existence seem to be limited \cite{Complex_Tarafder_2012}.
Last, but not least, the strain-induced ferromagnetic insulating
state in lanthanum cobaltite (LaCoO$_3$) thin film and its possible
relation with IS Co$^{3+}$ has attracted significant attention and
is still being debated \cite{LaCoO3_thinfilm_Choi_2012, Freeland
2008, Fuchs 2007, Fuchs 2008, Fuchs 2009,
LaCoO3_thinfilm_Fujioka_2013, Gupta 2008, Herklotz 2009,
LaCoO3_EFG_Hsu, LaCoO3_thinfilm, Mehta 2009, Merz 2010, Park 2009,
Pinta 2008, Posadas 2011, Rondinelli 2009}. A comprehensive
theoretical study of IS Fe$^{2+}$ in minerals with different atomic
structures under increasing pressure can be expected to provide
different perspectives for these problems.

\section{Computation}

In this work, all major calculations are performed using the local
density approximation + self-consistent Hubbard $U$ (LDA+$U_{sc}$)
method, as LDA+$U_{sc}$ gives the most accurate equation of states for iron-bearing Earth minerals and best predicts the occurence and mechanism of spin crossover in them, compared with other functionals \cite{QS MgFeSiO3 Hsu, Ferric, Ppv QS by Yu}.
Structural optimizations for 64-atom (Mg$_{1-x}$,Fe$_x$)O
supercell ($x=0.03125$) and 40-atom (Mg$_{1-x}$,Fe$_x$)SiO$_3$
supercell ($x=0.125$) in all possible spin states are performed with
variable cell shape molecular dynamics \cite{VC-relax} implemented
in the \textsc{quantum espresso} code \cite{PWscf}, which adopts the
plane-wave pseudopotential method. Pseudopotentials used in this
paper have been reported in Ref.~\onlinecite{Koichiro} and used in
other works regarding Earth minerals \cite{QS MgFeSiO3 Hsu, Ferric,
Fe-Al Ppv, Koichiro, Ppv QS by Yu}. A $4 \times 4 \times 4$
\textbf{k}-point mesh is used for both Fp and Fe-Pv supercells. In
this paper, we compute the Hubbard $U$ for each spin state with a
self-consistent procedure \cite{Campo10, DFT+U+J, Ferric, Usc by
Kulik}; the resultant Hubbard $U$ is referred to as self-consistent
$U$ ($U_{sc}$) hereafter. A detailed description of this procedure
can be found in Ref.~\onlinecite{Ferric} and its online supplemental
material. In brief, we start with an LDA+$U$ calculation with a trial
$U$ ($U_{in}$) to get all possible spin states. By applying local
perturbations to the iron site in the LDA+$U_{in}$ ground state with
the Hubbard potential being held fixed, the second derivative of the
LDA energy with respect to the electron occupation at the iron site
can be obtained using a linear response theory \cite{Cococcioni
LDA+U}. This second derivative, $U_{out}$, will be used as $U_{in}$
in the next iteration. Such a procedure is repeated until self
consistency is achieved, namely, $U_{in}=U_{out} \equiv U_{sc}$.

\bigskip

The iron nuclear quadrupole splitting (QS), $\Delta E_Q$, of each
possible spin state are computed using
\begin{equation}
\Delta E_Q = \frac{eQ|V_{zz}|}{2}\sqrt{1+\frac{\eta^2}{3}},
\label{QS}
\end{equation}
where $e$ is electron charge, $V_{zz}$ is the electric field
gradient (EFG), $\eta$ is the asymmetry parameter (usually small),
and $Q$ is the $^{57}$Fe nuclear quadrupole moment, determined to be
$0.16$ barn (1 barn $=10^{-28}$ m$^2$) \cite{EFG using PAW}. The EFG
and asymmetry parameter are computed using the WIEN2k code
\cite{WIEN2k}, which adopts the augmented plane-wave plus local
orbitals (APW+lo) method \cite{APW+lo}. Given that $Q=0.16$ barn is
sometimes considered underestimated, we also use $Q=0.18$ barn to
compute the upper limit of QS.

\section{Results and Discussion}

\subsection{IS Fe$^{2+}$ in Ferropericlase}

\begin{figure}[pt]
\begin{center}
\includegraphics[
]{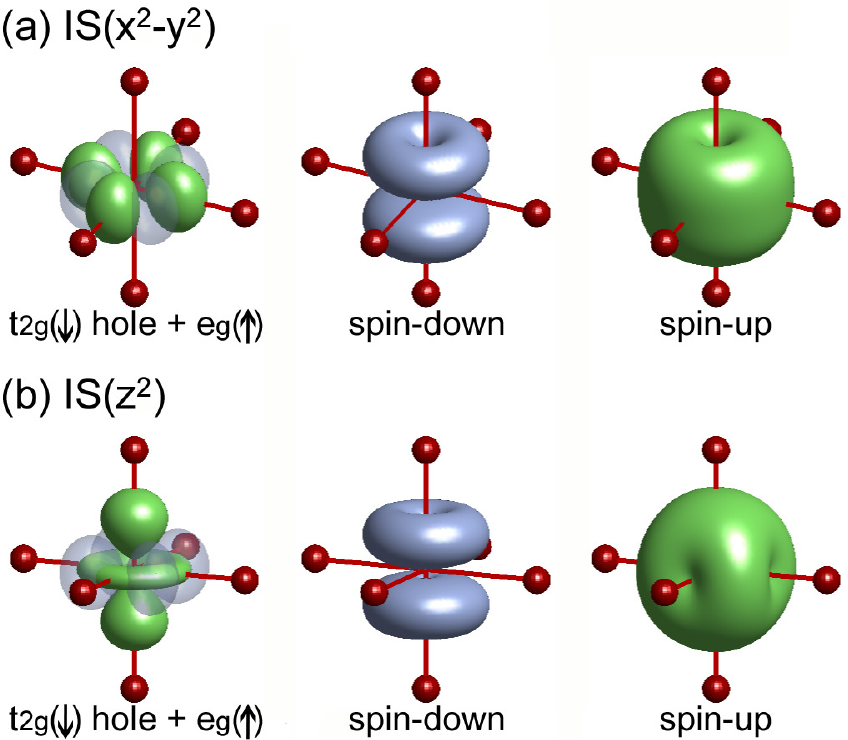}
\end{center}
\caption{Possible electronic configurations of an intermediate-spin
$d^{6}$ ion in a tetragonally distorted octahedral site. Schematic
plots of the combination of $t_{2g}$ hole (transparent blue) and
filled $e_g$ orbital (solid green), along with the spin-up/down
electron density are shown. (a) The IS($x^2-y^2$) state: a spin-down
$t_{2g}$ hole opened in the $d_{xy}$ orbital and the spin-up $e_g$
electron occupying the $d_{x^2-y^2}$ orbital; (b) The IS($z^2$)
state: a spin-down $d_{xy}$ hole and a spin-up $d_{z^2}$ electron.}
\label{3d6_IS_t2g_hole_eg}
\end{figure}

In Fp, Fe$^{2+}$ substitutes Mg$^{2+}$ in the MgO rock salt
structure, residing in an octahedral site surrounded by 6 oxygen
atoms. For a $d^6$ ion (e.g. Fe$^{2+}$, Co$^{3+}$,...) in such an
octahedral site, there can be only one HS ($t_{2g}^4e_g^2$) state
and one LS ($t_{2g}^6e_g^0$) state. The IS ($t_{2g}^5e_g^1$) state,
however, is not unique. A $t_{2g}^5e_g^1$ state has a spin-down hole
opened in the $t_{2g}$ manifold. By properly choosing the coordinate
system, this empty $t_{2g}$ orbital can always be $d_{xy}$. The one
filled spin-up $e_g$ orbital can be either $d_{x^2-y^2}$ or
$d_{z^2}$, forming two distinct IS states. Characterized by their
filled $e_g$ orbitals, these two IS states are referred to as the
IS($x^2-y^2$) and IS($z^2$) state, respectively, as shown in
Fig.~\ref{3d6_IS_t2g_hole_eg}. Since these two possible IS states
have different Jahn-Teller (J-T) active orbitals occupied, their J-T
distortions should be different as well: the IS($x^2-y^2$) state
would have elongated bond length on the $xy$ plane, while the
IS($z^2$) state would have elongated bond length along the $z$
direction.

\bigskip

\begin{figure}[pt]
\begin{center}
\includegraphics[
]{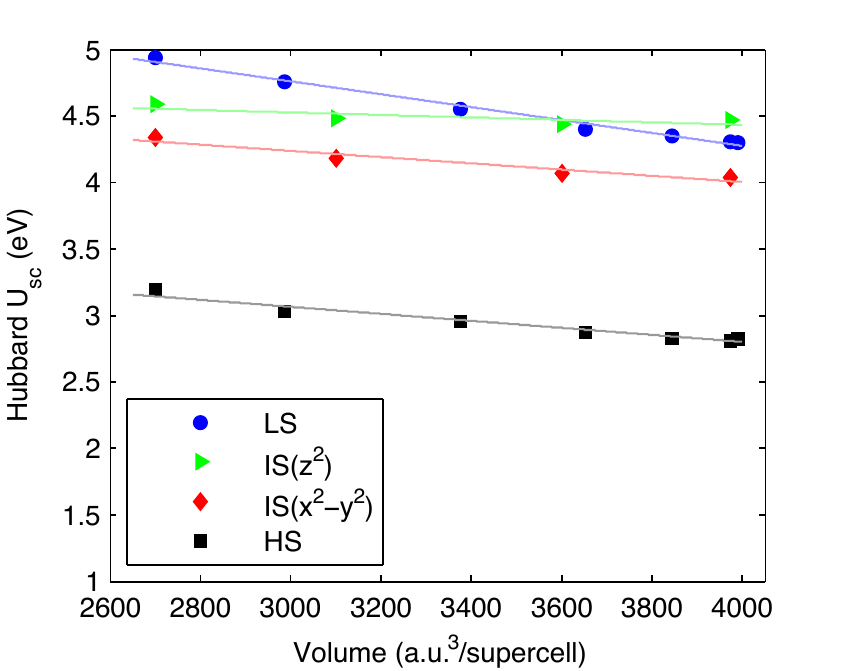}
\end{center}
\caption{The self-consistent $U$ ($U_{sc}$) of Fe$^{2+}$ in
(Mg$_{1-x}$,Fe$_x$)O ($x=0.03125$) at different 64-atom supercell
volumes ($V$) in the pressure range of 0--150 GPa (symbols). A
straight line provides an adequate fit for $U_{sc}(V)$.}
\label{Hubbard U}
\end{figure}

Using the LDA+$U$ method, all the above-mentioned spin states can be
obtained in Fp. The $U_{sc}$'s of these states at different 64-atom
Mg$_{1-x}$Fe$_x$O ($x=3.125$\%) supercell volumes are shown in
Fig.~\ref{Hubbard U}. In Fp, the $U_{sc}$ decreases with total
electron spin $S$, similar to Fe-Pv \cite{Ferric}, Fe-Ppv \cite{Ppv
QS by Yu,Fe-Al Ppv}, and LaCoO$_3$ \cite{LaCoO3_thinfilm}. Another
common feature shared by these systems is that the volume dependence
of $U_{sc}$ is marginal \cite{LS_LaCoO3_Hsu, Ferric,
LaCoO3_thinfilm, Fe-Al Ppv, Ppv QS by Yu}. Notably, the $U_{sc}$'s
of the two IS states are different, regardless of their same total
spin moment $S=1$. The $U_{sc}$ of the IS($z^2$) state is higher
than that of the IS($x^2-y^2$) state by 0.3--0.5 eV, indicating the
former has a stronger on-site Coulomb interaction. It should be
pointed out that the general trend of Hubbard $U$ decreasing with
total electron spin $S$ was not observed in an earlier calculation
on Fp \cite{MgFeO by Tsuchiya}. The main reason is that the
$U_{sc}$'s reported here are extracted from a series of trial
LDA+$U$ ground states, while the Hubbard $U$'s reported in
Ref.~\onlinecite{MgFeO by Tsuchiya} were extracted from the LDA
ground state. For the LS state in Fp, both LDA and LDA+$U$ methods
give an insulating ground state with the same orbital occupancy; the
Hubbard $U$ extracted from the LDA or LDA+$U$ ground states are thus
similar. For the HS state in Fp, however, LDA does not give a
correct orbital occupancy; it gives three partially (and equally)
occupied $t_{2g}$ orbitals by one spin-down electron, resulting in a
metallic state, in contrast with the LDA+$U$ insulating ground state
with one fully occupied $t_{2g}$ orbital by one spin-down electron.
The Hubbard $U$ of HS Fe$^{2+}$ in Fp reported in
Ref.~\onlinecite{MgFeO by Tsuchiya} is thus significantly different
from the $U_{sc}$ reported here.

\bigskip

\begin{figure}[pt]
\begin{center}
\includegraphics[
]{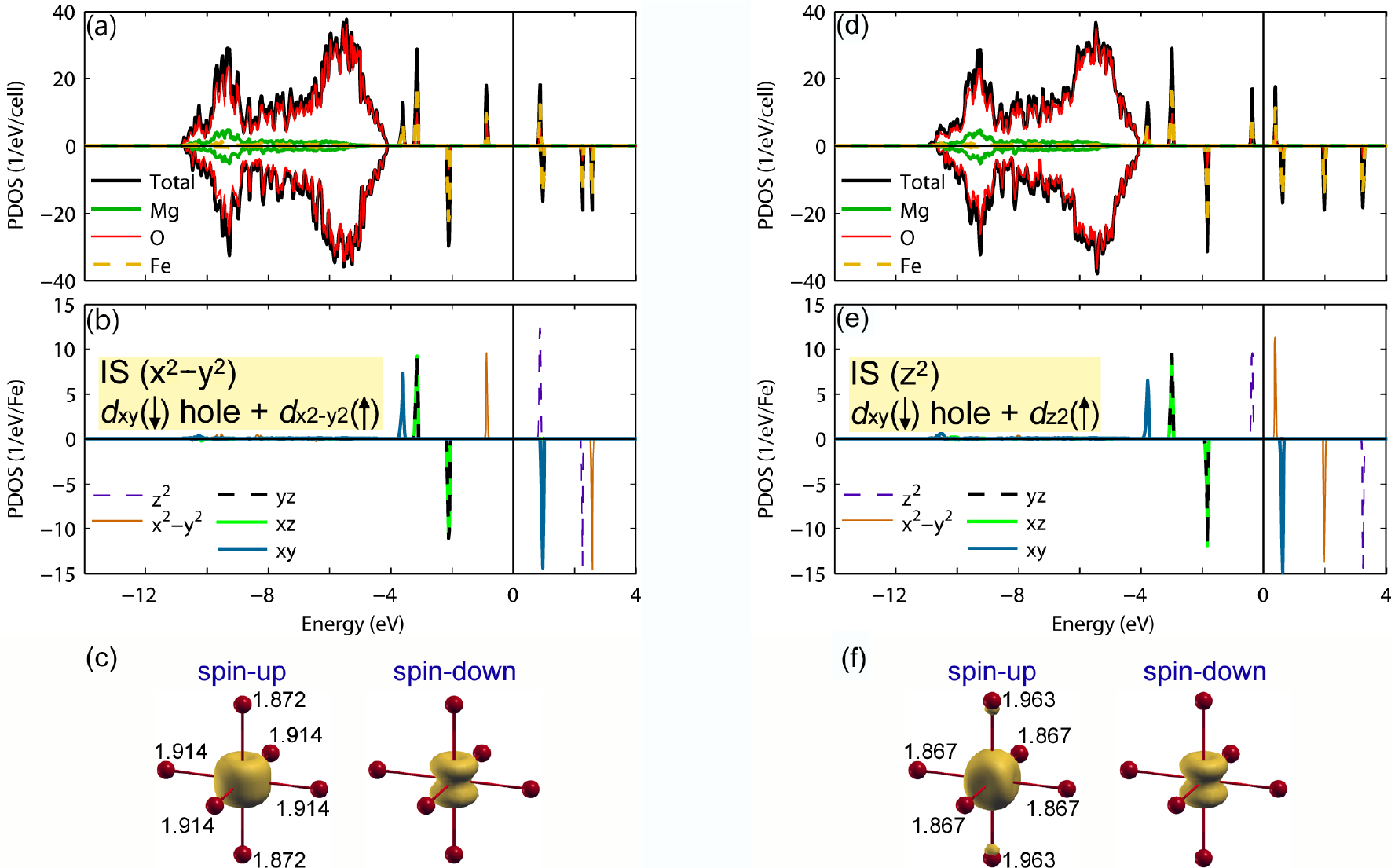}
\end{center}
\caption{Electronic structures of the IS($x^2-y^2$) state (a)-(c)
and the IS($z^2$) states (d)-(f) in (Mg$_{1-x}$,Fe$_x$)O
($x=0.03125$) at 112 GPa. (a),(d) PDOS decomposed by atomic species.
(b),(e) PDOS decomposed by cubic harmonic. (c),(f) The FeO$_6$
octahedron and ILDOS of Fe peaks shown in panels (a) and (d).
Numbers in panels (c) and (f) indicate the Fe--O distances (in
{\AA}).} \label{Fp_IS_electronic}
\end{figure}

The electronic structures and local Fe--O configurations of the two
IS states at 112 GPa are shown in Fig.~\ref{Fp_IS_electronic}. The
projected density of states (PDOS) of each cubic harmonic clearly
shows that the IS($x^2-y^2$) state has a filled spin-up
$d_{x^2-y^2}$ orbital [Fig.~\ref{Fp_IS_electronic}(b)], and the
IS($z^2$) state has a filled spin-up $d_{z^2}$ orbital
[Fig.~\ref{Fp_IS_electronic}(e)]. The integrated local density of
states (ILDOS) of all the Fe peaks in
Figs.~\ref{Fp_IS_electronic}(a) and \ref{Fp_IS_electronic}(d) are
plotted in Figs.~\ref{Fp_IS_electronic}(c) and
\ref{Fp_IS_electronic}(f), respectively. These calculation results
are consistent with the schematic plots shown in
Fig.~\ref{3d6_IS_t2g_hole_eg}. By comparing
Figs.~\ref{3d6_IS_t2g_hole_eg}(a) and \ref{Fp_IS_electronic}(c), the
IS($x^2-y^2$) state shows a clear $t_{2g}$+$d_{x^2-y^2}$ character
in its spin-up channel (an empty $d_{z^2}$ orbital can be observed);
by comparing Figs.~\ref{3d6_IS_t2g_hole_eg}(b) and
\ref{Fp_IS_electronic}(f), the IS($z^2$) state shows a clear
$t_{2g}$+$d_{z^2}$ character (an empty $d_{x^2-y^2}$ orbital can be
observed). As expected, these two IS states exhibit distinct J-T
distortions: the IS($x^2-y^2$) state has elongated Fe--O distances
on the $xy$ plane [Fig.~\ref{Fp_IS_electronic}(c)], while the
IS($z^2$) state has elongated Fe--O distances along the $z$ axis
[Fig.~\ref{Fp_IS_electronic}(f)]. 

\bigskip

With the orbital occupancy and charge density shown above, the
$U_{sc}$'s difference between the two IS states in Fp can be
qualitatively explained. Indeed, our discussion is based on Kohn-Sham orbitals, which are usually considered of little physical meaning. In practice, however, they
resemble the real electronic structure of most systems and can be used for molecular orbital or chemical analysis \cite{Stowasser_1999_JACS}. For the IS($z^2$) state, its filled $e_g$
orbital ($d_{z^2}$) is oriented vertically with the spin-down
$t_{2g}$ hole ($d_{xy}$) and \textit{passing through} the
donut-shape lobes of the spin-down electron charge density, as can
be observed in Fig.~\ref{3d6_IS_t2g_hole_eg}(b). In contrast, the
IS($x^2-y^2$) state has its filled $e_g$ orbital ($d_{x^2-y^2}$)
oriented on the same plane with the spin-down $t_{2g}$ hole,
\textit{laying in between} the donut-shape lobes of the spin-down
electrons, as can be observed in Fig.~\ref{3d6_IS_t2g_hole_eg}(a).
Clearly, the $e_{g}$ electron of the IS($z^2$) state overlaps with
the spin-down electrons more than that in the IS($x^2-y^2$) state.
This greater electron-electron overlap in the IS($z^2$) state leads
to a stronger on-site Coulomb interaction and thus a higher
$U_{sc}$. 

\bigskip

\begin{figure}[pt]
\begin{center}
\includegraphics[
]{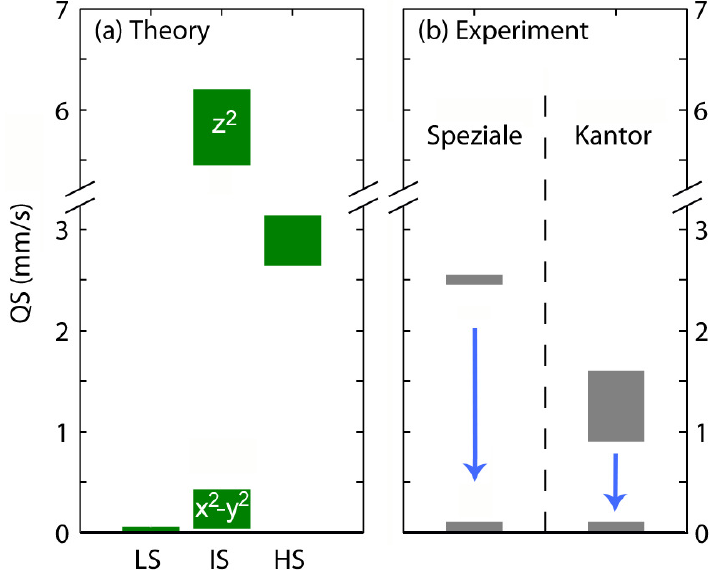}
\end{center}
\caption{Calculated iron nuclear QS in ferropericlase (a) and
experimental values (b) by Speziale \textit{at al}. \cite{MgFeO by
Speziale 2005} and Kantor \textit{et al}. \cite{MgFeO by Kantor
2009}. Arrows in panel (b) indicate the drastic change in QS with
increasing pressure.} \label{Fp_QS}
\end{figure}

One reliable way to identify iron spin state in Earth minerals is by
comparing the iron nuclear QS obtained by theory and experiments, as
demonstrated previously in Fe-Pv/Ppv \cite{QS MgFeSiO3 Hsu, Ferric,
Fe-Al Ppv, Ppv QS by Yu}. The same approach can be applied to Fp as
well. For each spin state, we compute the lower and upper limit of
iron nuclear QS in the lower-mantle pressure range (see Section 2).
The calculation results are shown in Fig.~\ref{Fp_QS}(a), along with
the measured QS (via M\"{o}ssbauer spectroscopy) by Speziale
\textit{et al}. \cite{MgFeO by Speziale 2005} and Kantor \textit{et
al}. \cite{MgFeO by Kantor 2009} shown in Fig.~\ref{Fp_QS}(b). The
dependence of QS on spin state can be understood via the electric
field gradient, $V_{zz}$, at the iron nucleus. The QS is directly
proportional to the EFG magnitude ($|V_{zz}|$), as shown in
Eq.~\ref{QS}, and the $d$ electrons contribute to the EFG as the
following:
\begin{equation}
V_{zz}\propto\sum_{\sigma} \left(
2n_{x^{2}-y^{2}}^{\sigma}-2n_{z^{2}}
^{\sigma}+2n_{xy}^{\sigma}-n_{yz}^{\sigma}-n_{xz}^{\sigma}\right)
/\langle r^{3}\rangle, \label{Vzz}
\end{equation}
where $n^{\sigma}_{xy}$, $n^{\sigma}_{yz}$,... are the occupancy of
each $3d$ orbital by the electron with spin $\sigma$ [up
($\uparrow$) or down ($\downarrow$)]. Clearly, the dependence of QS
on iron spin state results from the dependence of EFG on orbital
occupancy, therefore, iron spin state cannot be directly derived
from the numerical value of QS. In Fp, LS Fe$^{2+}$ resides in an
octahedral site with cubic ($O_h$) symmetry and has $n^{\sigma}_{xy}
= n^{\sigma}_{yz} = n^{\sigma}_{xz} \approx 1$ and
$n^{\sigma}_{x^2-y^2} = n^{\sigma}_{z^2} \approx 0$ for both spin up
($\sigma=\uparrow$) and spin down ($\sigma=\downarrow$). Its QS,
based on Eq.~\ref{Vzz}, should be zero, consistent with
Fig.~\ref{Fp_QS}(a). The HS Fe$^{2+}$ has five spin-up electrons
($\sigma=\uparrow$) occupying all $3d$ orbitals forming a spherical
shaped charge density. Evident from Eq.~\ref{Vzz}, these spin-up
electrons barely contribute to the EFG. It is the one spin-down
($\sigma=\downarrow$) electron, $d_{xy}$, that contributes to the
EFG. The computed QS of the HS state is about 2.6--3.1 mm/s, similar
to that reported in Ref.~\onlinecite{MgFeO by Speziale 2005}, but
higher than that reported in Ref.~\onlinecite{MgFeO by Kantor 2009}
[Fig.~\ref{Fp_QS}(b)]. The IS($x^2-y^2$) state has
$n^{\uparrow}_{x^2-y^2} \approx n^{\uparrow}_{xy} \approx
n^{\uparrow}_{yz} \approx n^{\uparrow}_{xz} \approx 1$,
$n^{\downarrow}_{xz} \approx n^{\downarrow}_{yz} \approx 1$, and the
remaining orbitals being empty. Such a configuration, based on
Eq.~\ref{Vzz}, would lead to an almost vanishing EFG and thus a very
small QS. In contrast, the IS($z^2$) state has $n^{\uparrow}_{z^2}
\approx n^{\uparrow}_{xy} \approx n^{\uparrow}_{yz} \approx
n^{\uparrow}_{xz} \approx 1$ and $n^{\downarrow}_{xz} \approx
n^{\downarrow}_{yz} \approx 1$. This would lead to an EFG twice
larger (in magnitude) than that of the HS state. Indeed, the
computed QS of the IS($z^2$) state is 5.5--6.2 mm/s. Such an
exceptionally high QS is not observed in Fp. In this sense, the
possibility of HS-IS($z^2$)-LS crossover can be ruled out. While the
remaining possible scenarios, HS-LS and HS-IS($x^2-y^2$)-LS
crossovers, are both consistent with M\"{o}ssbauer spectra,
first-principles calculations can provide further information to pin
down the spin-crossover mechanism, as described below.

\bigskip

\begin{figure}[pt]
\begin{center}
\includegraphics[
]{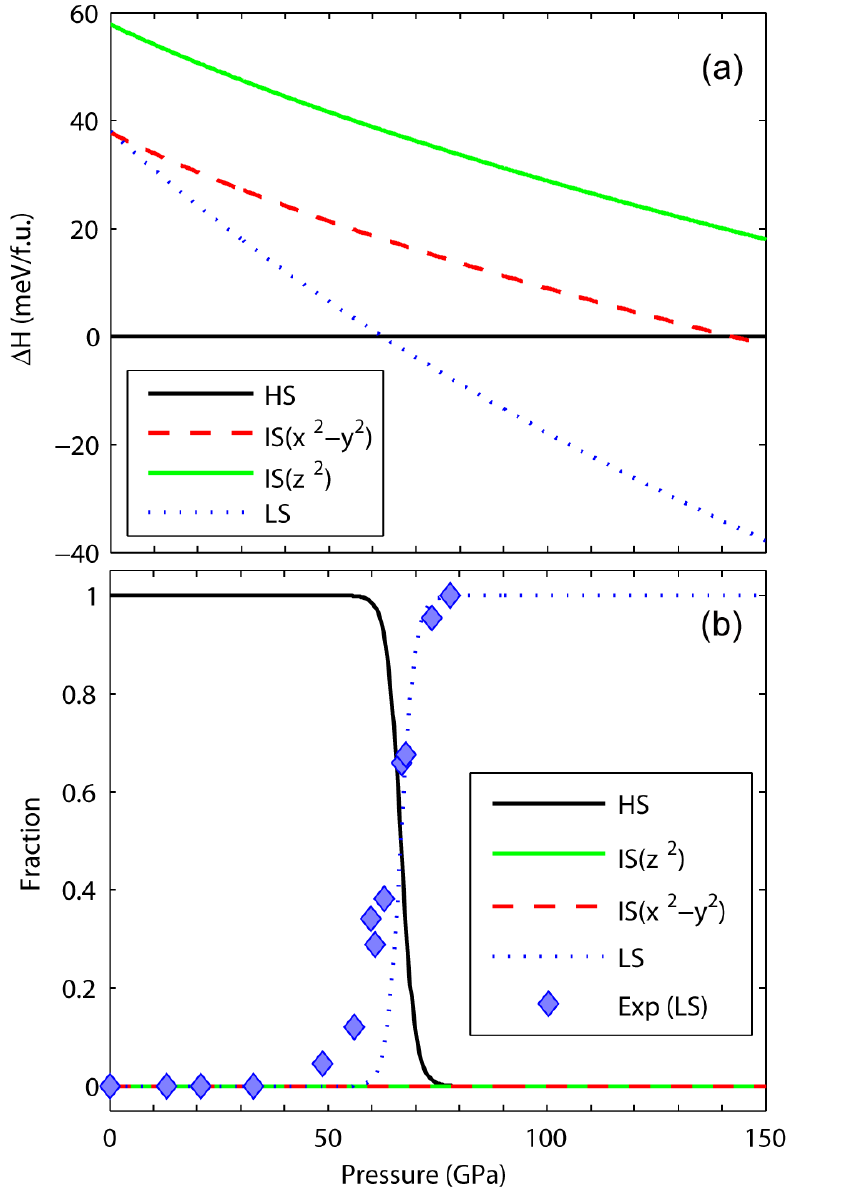}
\end{center}
\caption{(a) Relative enthalpy ($\Delta H$) of (Mg$_{1-x}$,Fe$_x$)O
($x=0.03125$) in each spin state with respect to the HS state. (b)
Molar fraction of each spin state at room temperature predicted by
theory (lines) and the LS fraction extracted from the M\"{o}ssbauer
spectra ($x=0.05$ in the sample) \cite{MgFeO by Kantor 2009}. }
\label{Fp_dH_n}
\end{figure}

Using the LDA+$U_{sc}$ method, the equation of states and energetics
of (Mg$_{1-x}$,Fe$_x$)O ($x=3.125\%$) in all spin states can be
computed; the relative enthalpies ($\Delta H_{i}$) of each spin
state $i$ [$i=$ HS, IS($x^2-y^2$), IS($z^2$), or LS] with respect to
the HS state are plotted in Fig.~\ref{Fp_dH_n}(a). With known
$\Delta H_{i}$, the fraction ($n_i$) of each spin state can be
estimated using the following expression derived from a
thermodynamic model detailed in Refs.~\onlinecite{Site degeneracy
Umemoto} and \onlinecite{Fp Pv by Hsu}, subject to the constraint
$\sum_i{n_i}=1$,
\begin{equation}
n_i(P,T)=n_{HS} \times \frac{m_i(2S_i+1)}{m_{HS}(2S_{HS}+1)} \times
\exp(-\frac{\Delta H_i}{k_BTx}) ~\text{for} ~i\neq ~\text{HS},
\label{ni}
\end{equation}
where $m_i$ and $S_i$ are the orbital degeneracy and total spin
moment of spin state $i$, respectively. In Fp, $m_{HS}=m_{IS}=3$
(for both types of IS), and $m_{LS}=1$. The fraction $n_i$ of each
spin state at room temperature ($T=300$ K) are plotted in
Fig.~\ref{Fp_dH_n}(b). Here, we do not include vibrational free
energy, as it only slightly increases the transition pressure
\cite{Wentzcovitch PNAS, MgFeO by Wu} and would not change the main
conclusion: populations of the IS states are too low to be observed
due to their extremely high enthalpies. This result is consistent
with the lack of a QS of 5.5--6.2 mm/s in M\"{o}ssbauer spectra; it
also confirms the small QS observed in Fp should be the LS, not the
IS($x^2-y^2$) state, showing that Fp undergoes a HS-LS crossover. We
can also observe in Fig.~\ref{Fp_dH_n}(b) that overall, the computed
LS fraction agrees very well with that derived from M\"{o}ssbauer
spectra of a sample with iron concentration $x=0.05$ \cite{MgFeO by
Kantor 2009}. The small discrepancy is that in experiment, the LS
fraction ($n_{LS}$) reaches 10\% at $\sim 55$ GPa, while our
calculation predicts $\sim 62$ GPa. Indeed, the transition pressure
predicted by LDA+$U_{sc}$ is slightly higher than that observed
Ref.~\onlinecite{MgFeO by Kantor 2009} and other works
comprehensively reviewed in Ref.~\onlinecite{Fp Pv by Lin 2013}.
Such a discrepancy may be better addressed by including the exchange
term $J$ computed self-consistently \cite{DFT+U+J}. The HS state has
a larger $J$ than the LS state, which would increase the enthalpy of
the HS state more, lower the relative enthalpy of the LS state
($\Delta H_{LS}$), and thus lower the transition pressure.

\bigskip

While both IS states in Fp are unfavorable, a further analysis of
this simple system can help us better understand IS Fe$^{2+}$ in
more complicated environments, including (Mg,Fe)SiO$_3$ Pv and Ppv.
For Fe$^{2+}$ (or any $d^6$ ion) in an octahedral site, an IS state
can be produced from the LS state by opening a spin-down $t_{2g}$
hole and filling a spin-up electron in an $e_g$ orbital. We have
shown that there can be more than one possible combination of
$t_{2g}$ hole and $e_g$ electron (referred to as hole-electron
combination hereafter). Furthermore, we shall see that among the
possible IS states, the one with a closely oriented hole-electron
combination has lower enthalpy, making it the \textit{most probable}
IS state. In Fp, the IS($x^2-y^2$) state is the most probable IS
state. Its filled $e_g$ orbital, $d_{x^2-y^2}$, is closely oriented
with its $t_{2g}$ hole, $d_{xy}$ ($d_{x^2-y^2}$ is simply a rotation
with respect to $d_{xy}$ about the $z$-axis). Such a combination
leads to a smaller overlap between the spin-up $e_g$ electron and
spin-down electrons and thus leads to a less strong on-site Coulomb
interaction, smaller $U_{sc}$ (as described previously), and a lower
total energy. In contrast, the filled $e_g$ orbital of the IS($z^2$)
state, $d_{z^2}$, is vertically oriented with the $t_{2g}$ hole,
$d_{xy}$. The overlap between its spin-up $e_g$ electron and
spin-down electrons, the on-site Coulomb interaction, and the
totally energy are all larger.

\bigskip

Another attribute of the most probable IS Fe$^{2+}$ is its lower QS
compared with other IS sates. This is also a consequence of the
closely orientated hole and electron. For example, the IS($x^2-y^2$)
state has a $d_{x^2-y^2}$ electron and a $d_{xy}$ hole. Evident from
Eq.~\ref{Vzz}, $d_{x^2-y^2}$ and $d_{xy}$ contribute equally to
$V_{zz}$. Since the $d_{x^2-y^2}$ orbital is produced by a rotation
of the $d_{xy}$ orbital about the $z$-axis, they should both have
the same second derivative along the $z$ direction. Starting with an
LS state, opening a $d_{xy}$ hole followed by filling a
$d_{x^2-y^2}$ electron would not significantly change the EFG.
Therefore, the QS of IS($x^2-y^2$) state should be very similar to
that of the LS state. In contrast, the IS($z^2$) is configured by
opening a hole in $d_{xy}$ of a LS state, followed by filling an
electron in a $d_{z^2}$, vertically oriented to $d_{xy}$. This would
severely change the EFG and lead to a very different QS from the LS
state. Based on this analysis, an IS Fe$^{2+}$(or $d^6$ ion) in a
more complicated crystal-field environment could still have its
energy lowered by bringing the hole-electron combination to a close
configuration as in $d_{xy}$-$d_{x^2-y^2}$ that leads to a low QS.

\subsection{IS Fe$^{2+}$ in MgSiO$_3$ perovskite and post-perovskite }

\begin{figure}[pt]
\begin{center}
\includegraphics[
]{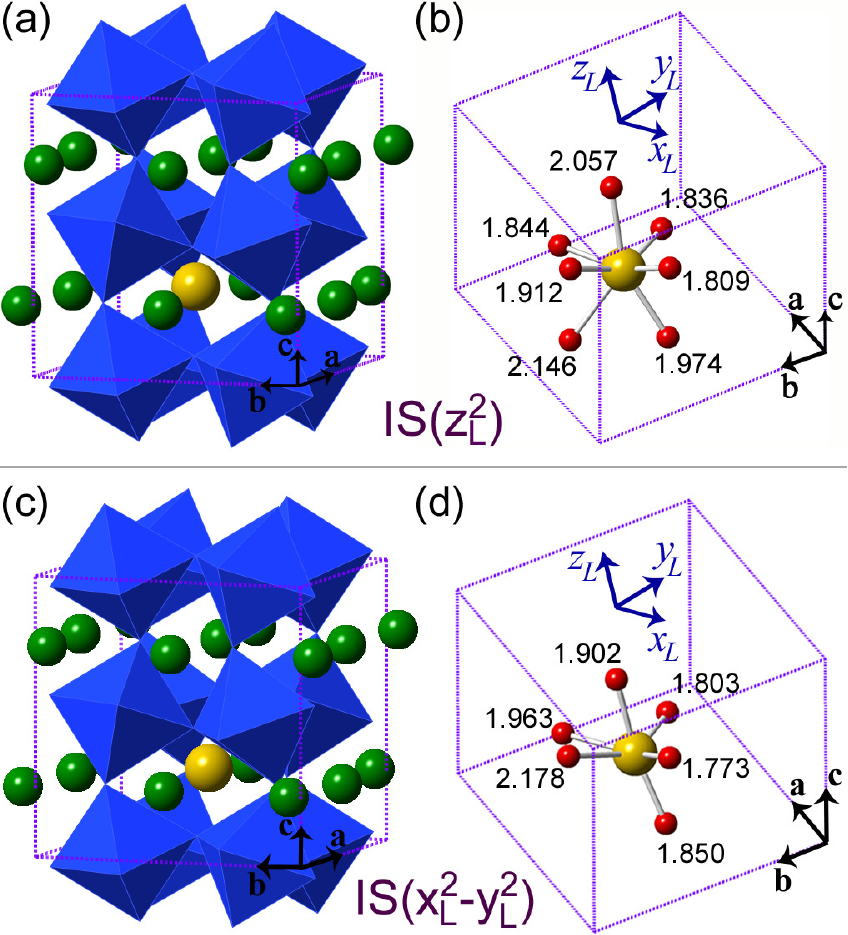}
\end{center}
\caption{Atomic structure (a),(c) and the local Fe--O configuration
(b),(d) of (Mg$_{1-x}$,Fe$_{x}$)SiO$_3$ perovskite ($x=0.125$) with
IS($z_L^2$) and IS($x_L^2-y_L^2$) Fe$^{2+}$ at 120 GPa. The large
(yellow), medium (green), and small (red) spheres are Fe, Mg, and O
atoms, respectively; the octahedra (blue) are SiO$_6$ octahedra. The
dotted (purple) line indicates the 40-atom super cell. Numbers in
panels (b) and (d) indicate the Fe--O distances (in {\AA}). A set of
Fe--O local coordinate ($x_L$, $y_L$, $z_L$) can be defined, and
they do not align with the crystallographic coordinate (\textbf{a},
\textbf{b}, \textbf{c}).} \label{Pv_Ferrous_IS_atomic}
\end{figure}

As mentioned in Section I, first-principles computations so far do
not support HS-IS crossover of Fe$^{2+}$ in (Mg,Fe)SiO$_3$ Pv but
point to a crossover between two HS states with distinct QSs instead
\cite{QS MgFeSiO3 Hsu, Ferric}. The finding of two distinct types of
IS Fe$^{2+}$ in Fp, however, suggests that further investigations,
in particular, a thorough search for IS Fe$^{2+}$ in Pv, would be
necessary. Such a search, however, is not as straightforward as in
Fp, as the atomic structure of Pv is more complicated, and the
orbital occupancies of IS Fe$^{2+}$ in Pv are not known \textit{a
priori}. To make sure all possible orbital occupancies are
investigated, we produce IS Fe$^{2+}$ by manipulating the orbital
occupancy of LS Fe$^{2+}$. The reason is that LS and IS Fe$^{2+}$
both displace from the high-symmetry mirror plane (in contrast to HS
Fe$^{2+}$) to a position where only 6 oxygens are close enough to
significantly affect the iron $3d$ electrons, namely, both LS and IS
Fe$^{2+}$ reside in a highly distorted octahedral crystal site
\cite{Site degeneracy Umemoto}. As a consequence, the LS Fe$^{2+}$
has three doubly occupied $t_{2g}$-like and two empty $e_{g}$-like
orbitals \cite{Koichiro}. An IS state can thus be produced by
opening a spin-down hole in a $t_{2g}$-like orbital, filling a
spin-up electron in an $e_g$-like orbital, followed by a structural
optimization. Given the lack of symmetry in the distorted octahedral
site, there are six possible hole-electron combinations to be tested
(three inequivalent $t_{2g}$-like holes and two $e_g$-like
orbitals), in contrast to two in Fp, where only tetragonal
distortions are allowed. Among these six hole-electron combinations,
only two can be stabilized. Characterized by their filled
$e_{g}$-like orbitals, these two states are referred to as the
IS($z_L^2$) and IS($x_L^2-y_L^2$) states, respectively, and they
both have a $U_{sc}$ of 4.3 eV. As shall be detailed below, however,
these IS states are not exactly the same as those in Fp. With
different orbital occupancies, the position of these two IS
Fe$^{2+}$ in the big cage and the local Fe--O configurations are
different. Their atomic structures at 120 GPa are shown in
Fig.~\ref{Pv_Ferrous_IS_atomic}, where the numbers in panels (b) and
(d) are the Fe--O distances (in \AA). For each case, a local
coordinate system ($x_L$, $y_L$, $z_L$) based on the Fe--O bonds can
be defined, and it does not align with the crystallographic
coordinates (\textbf{a}, \textbf{b}, \textbf{c}). In such a highly
distorted octahedral crystal field, the $d$ orbitals would no longer
be cubic harmonics.

\bigskip

\begin{figure}[pt]
\begin{center}
\includegraphics[
]{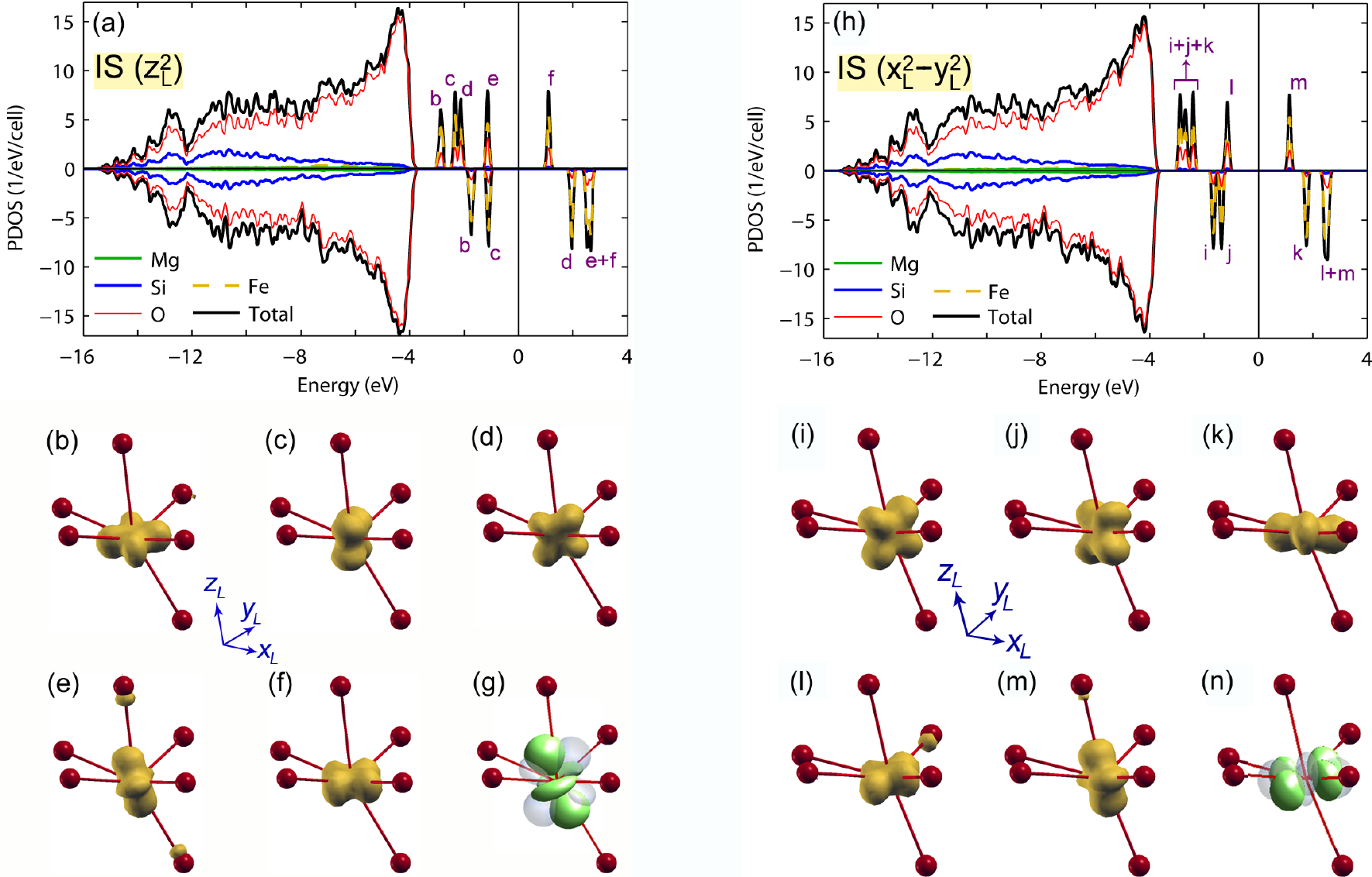}
\end{center}
\caption{Electronic structure of IS($z_L^2$) state (a)-(g) and
IS($x_L^2-y_L^2$) state (h)-(n) of (Mg$_{1-x}$,Fe$_{x}$)SiO$_3$
perovskite ($x=0.125$) at 120 GPa. (a),(h) PDOS decomposed by atomic
species, where peaks b-f and i-m are contributed by iron, and each
of their ILDOS are plotted in panels (b)-(f) and (i)-(m)
respectively. The Fe--O distorted octahedra are the same as in
Figs.~\ref{Pv_Ferrous_IS_atomic}(b) and
\ref{Pv_Ferrous_IS_atomic}(d), with the longest Fe--O (2.146 \AA) in
\ref{Pv_Ferrous_IS_atomic}(b) omitted. Panels (b)-(d) and (i)-(k)
exhibit $t_{2g}$ character, while panels (e)-(f) and (l)-(m) exhibit
$e_{g}$ character. The IS($z_L^2$) state has a $t_{2g}$-like hole
opened in the $\sim (d_{x_Lz_L}+d_{y_Lz_L})/\sqrt{2}$ orbital (d)
and an $e_{g}$-like electron occupying the $\sim d_{z_L^2}$ orbital
(e). The IS($x_L^2-x_L^2$) state has a $t_{2g}$-like hole opened in
the $\sim (d_{x_Ly_L})$ orbital (k) and an $e_{g}$-like electron
occupying the $\sim d_{x_L^2-y_L^2}$ orbital (l). The hole-electron
combination of these two states are schematically depicted in panels
(g) and (n), where the transparent (gray) and solid (green) surfaces
indicate the hole and the electron, respectively.}
\label{Pv_Ferrous_IS_electronic}
\end{figure}

Among the two IS states, the IS($z_L^2$) state is briefly reported
in Ref.~\onlinecite{QS MgFeSiO3 Hsu} without insightful analysis.
The PDOS of this state at 120 GPa is shown in
Fig.~\ref{Pv_Ferrous_IS_electronic}(a), where the peaks (indicated
by letters b-f) resulting from iron $3d$ electrons can be clearly
observed. The ILDOS of each peak are plotted in
Figs.~\ref{Pv_Ferrous_IS_electronic}(b)-\ref{Pv_Ferrous_IS_electronic}(f),
respectively. The lowest three orbitals (b, c, and d) exhibit
$t_{2g}$ character. In terms of the locally defined Fe--O coordinate
($x_L$, $y_L$, $z_L$), these three orbitals are $\sim d_{x_Ly_L}$,
$\sim (d_{x_Lz_L}-d_{y_Lz_L})/\sqrt{2}$, and $\sim
(d_{x_Lz_L}+d_{y_Lz_L})/\sqrt{2}$, respectively. The other two
orbitals (e and f) exhibit $e_{g}$ characters; they are $\sim
d_{z_L^2}$ and $\sim d_{x_L^2-y_L^2}$, respectively. The filled
$e_g$-like orbital, $\sim d_{z_L^2}$, is consistent with the longer
Fe--O distance in the $z_L$ direction
[Fig.~\ref{Pv_Ferrous_IS_atomic}(b)]. In this state, the spin-down
hole is opened in the $\sim (d_{x_Lz_L}+d_{y_Lz_L})/\sqrt{2}$
orbital, closely oriented with the $\sim d_{z_L^2}$ orbital. Such a
hole-electron combination is schematically depicted in
Fig.~\ref{Pv_Ferrous_IS_electronic}(g). As can be observed, the
filled $e_{g}$-like orbital of this state is almost a rotation with
respect to the $t_{2g}$-like hole. In this sense, this state is more
similar to the IS($x^2-y^2$) state in Fp, rather than the IS($z^2$)
in Fp. Therefore, the QS of this state is quite low, about 0.9--1.6
mm/s (depending on pressure), slightly higher than the QS of LS
Fe$^{2+}$ in Pv, 0.8 mm/s \cite{QS MgFeSiO3 Hsu, Ferric}. The PDOS
of IS($x_L^2-y_L^2$) state is shown in
Fig.~\ref{Pv_Ferrous_IS_electronic}(h). The ILDOS of each peak (i-m)
resulting from iron $3d$ electrons are shown in
Figs.~\ref{Pv_Ferrous_IS_electronic}(i)-\ref{Pv_Ferrous_IS_electronic}(m),
respectively. The lowest three orbitals (i, j, and k) exhibit
$t_{2g}$ characters: $\sim d_{y_Lz_L}$, $\sim d_{x_Lz_L}$, and $\sim
d_{x_Ly_L}$, respectively. The filled $e_g$-like orbital is $\sim
d_{x^2-y^2}$, consistent with the longer average Fe--O distances on
the $x_Ly_L$ plane [Fig.~\ref{Pv_Ferrous_IS_atomic}(d)]. This state
has a hole in the $\sim d_{x_Ly_L}$ orbital; it resembles the
IS($x^2-y^2$) state in Fp. As expected, it has quite low QS,
0.8--1.4 mm/s (depending on pressure), which is in between the QSs
of the LS and IS($z^2$) state in Pv.

\bigskip

\begin{figure}[pt]
\begin{center}
\includegraphics[
]{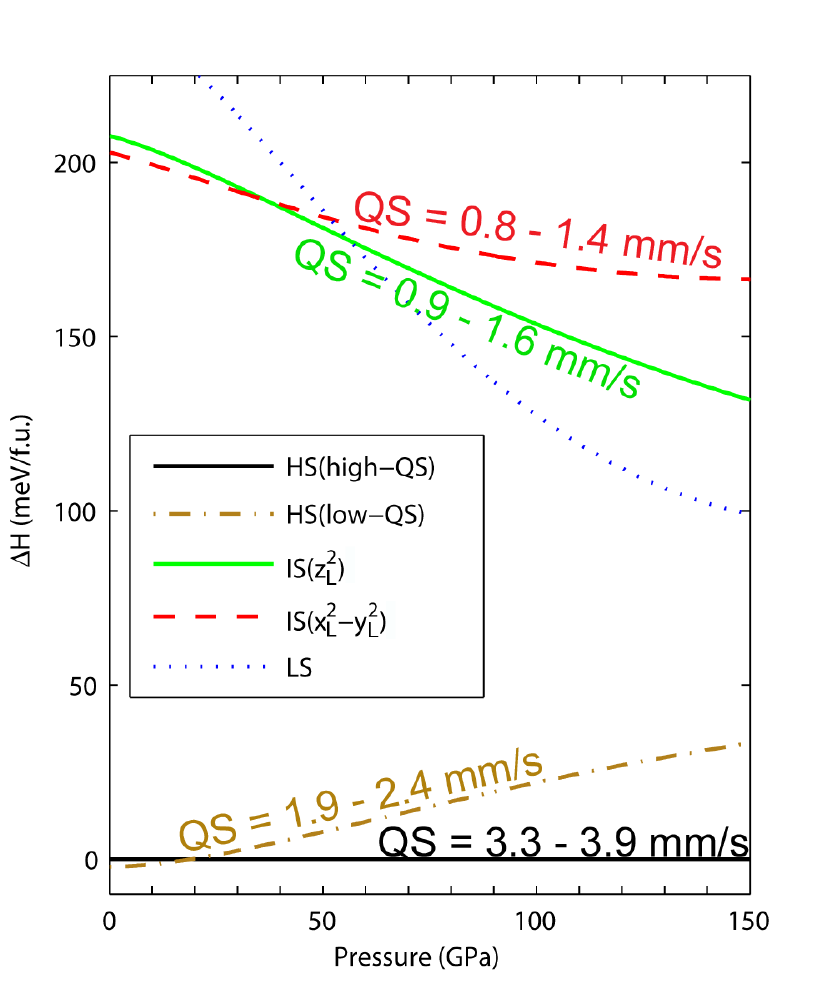}
\end{center}
\caption{Relative enthalpy ($\Delta H$) of
(Mg$_{1-x}$,Fe$_{x}$)SiO$_3$ perovskite ($x=0.125$) in each spin
state with respect to the high-QS HS state. The QSs of HS states
adopted from Ref.~\onlinecite{Ferric}. As indicated, a QS of 3.5
mm/s is \textit{not} a signature of IS Fe$^{2+}$.}
\label{dH_vs_P_all_Pv_PRB}
\end{figure}

To determine whether IS Fe$^{2+}$ in Pv is possible at all, we
compute the enthalpy of all possible spin states, including the two
HS states with distinct QSs (referred to as low-QS and high-QS HS
states) reported in Refs.~\onlinecite{QS MgFeSiO3 Hsu} and
\onlinecite{Ferric}, the two IS states, and one LS state. The
relative enthalpy of these states with respect to the high-QS HS
state, along with the computed QSs, are shown in
Fig.~\ref{dH_vs_P_all_Pv_PRB} (QSs of HS states are adopted from
Ref.~\onlinecite{Ferric}). The $U_{sc}$ of the HS and LS Fe$^{2+}$
are 3.1 and 4.5 eV, respectively, nearly the same as those reported
in Ref.~\onlinecite{QS MgFeSiO3 Hsu}, which extracts $U$ from the
DFT ground states. The reason is that for (Mg,Fe)SiO$_3$ Pv with low
iron concentration, standard DFT functionals can give correct
insulating state and orbital occupancy for the HS and LS states;
extracting $U$ from DFT or DFT+$U$ ground states should thus give
similar results. Also, the $U_{sc}$ of Fe$^{2+}$ in Pv barely
depends on supercell volume and can be treated as a constant with
respect to volume. Evident from Fig.~\ref{dH_vs_P_all_Pv_PRB}, the
two IS states are energetically competitive. In 0--32 GPa, the
IS($x_L^2-y_L^2$) state has lower enthalpy; in 32--150 GPa, which
covers most of the lower-mantle pressure range, the IS($z_L^2$) has
lower enthalpy, making it the most probable IS state in the lower
mantle. The reason why these two IS states have similar physical
properties (QS, $U_{sc}$, and enthalpy) is that the IS($z_L^2$)
state in Pv has its $t_{2g}$ hole in the
$(d_{x_Lz_L}+d_{y_Lz_L})/\sqrt{2}$ orbital, closely oriented with
the $d_{z_L}^2$ orbital, instead of in the $d_{x_Ly_L}$ orbital.
Hypothetically, if the $t_{2g}$-like hole of the IS($z_L^2$) state
in Pv were opened in the $d_{x_Ly_L}$ orbital like the IS($z^2$)
state in Fp, the on-site Coulomb interaction would be stronger, the
self-consistent Hubbard $U_{sc}$ would be larger, the total energy
would be significantly higher, and the QS would be exceptionally
large. To reduce the energy, both IS states in (Mg,Fe)SiO$_3$ Pv
have closely oriented hole-electron combination, similar to the
IS($x^2-y^2$) state in Fp. Regardless, these two IS states are still
not favorable. The only crossover in this system occurs between
low-QS to high-QS HS state at 20 GPa, similar to the previous
calculation \cite{QS MgFeSiO3 Hsu}.

\bigskip

As shown above, IS Fe$^{2+}$ in (Mg,Fe)SiO$_3$ Pv resides in a
distorted octahedral crystal field, with orbitals exhibiting
$t_{2g}$ and $e_g$ characters. With one $e_g$-like orbital being
filled, the $t_{2g}$-like hole is uniquely determined as well: it
should be opened in the most closely oriented orbital to reduce the
total energy. Given that there are only two $e_g$-like orbitals, the
two IS states reported here should include all possible IS Fe$^{2+}$
in Pv. The signature of IS Fe$^{2+}$ in Pv in the lower-mantle
pressure range should be a QS in between 0.8 and 1.6 mm/s. Given
the lack of such QS observed in M\"{o}ssbauer spectra and the high enthalpy of IS state, the
observed QS (3.5 mm/s \cite{QS-by-McCammon}) should be indeed a HS
state, and IS Fe$^{2+}$ in Pv would be highly unlikely.

\bigskip

While a similar investigation for Fe-Ppv is not conducted here, an
IS Fe$^{2+}$ in Ppv has been reported in Ref.~\onlinecite{Ppv QS by
Yu}, and it is similar to the IS($z_L^2$) state in Pv. Given the
highly similar crystal fields experienced by Fe$^{2+}$ in Pv and
Ppv, this reported IS state in Ppv should be the most probable IS
state in the D" pressure range, if not only. Nevertheless, this IS
state is still unfavorable, and its QS is inconsistent with
experiments either, as detailed in Ref.~\onlinecite{Ppv QS by Yu}.
Therefore, IS Fe$^{2+}$ in Ppv should also be highly unlikely.

\section{Conclusion}

Using LDA+$U_{sc}$ calculations, we have investigated in details the
possible stability of the controversial intermediate-spin state of
Fe$^{2+}$ in lower-mantle minerals subject to pressure-induced spin
crossover: ferropericlase [(Mg$_{1-x}$,Fe$_x$)O] ($x=0.03125$) and
(Mg$_{1-x}$,Fe$_x$)SiO$_3$ perovskite ($x=0.125$). Two types of IS
states with distinct $3d$ hole-electron combinations were found in
Fp: the IS($x^2-y^2$) state and the IS($z^2$) state, with a
$d_{x^2-y^2}$ and $d_{z^2}$ electron, respectively, and a $d_{xy}$
hole. These distinct orbital occupancies lead to distinct
Jahn-Teller distortions and iron nuclear quadrupole splittings: the
IS($z^2$) state has an exceptionally high QS ($\geq 5.5$ mm/s), and
the IS($x^2-y^2$) state has a quite low QS ($<0.5$ mm/s). The
on-site Coulomb interaction and the total energy of the
IS($x^2-y^2$) state are both lower than that of the IS($z^2$) state
because of its closely oriented hole-electron combination, namely,
less overlap between the spin-up $e_g$ and spin-down $t_{2g}$
electrons. In (Mg,Fe)SiO$_3$ Pv, although IS Fe$^{2+}$ resides in
the large dodecahedral site, it effectively experiences a distorted
octahedral crystal field. Two types of IS states are found, and they
can also be characterized by their filled $e_g$-like orbitals. The
hole-electron combination of these two IS states are both closely
oriented; they both exhibit characters similar to the IS($x^2-y^2$)
state in Fp. Therefore, these two IS Fe$^{2+}$ in Pv have similarly
low QS ($<1.6$ mm/s) and the same Hubbard $U_{sc}$, and they are
energetically competitive. Compared to the HS and LS states, all the
above-mentioned IS states in Fp and (Mg,Fe)SiO$_3$ Pv/Ppv are
energetically unfavorable; their QSs are also all inconsistent with
experiments. Most importantly, these considered IS Fe$^{2+}$ already
include all relevant types of IS Fe$^{2+}$ in lower-mantle minerals.
Therefore, it is highly unlikely that IS Fe$^{2+}$ exists in the
lower mantle.

\bigskip

Finally, although this present work is mainly focused on
lower-mantle minerals under pressure (variable metal-oxygen
distance), it is an exemplar of the behavior of other strongly
correlated $d^6$ ions in two common crystalline sites of complex
oxides: the octahedral ($B$) site in ABO$_3$ perovskite and in the
rocksalt structure, and the dodecohedral ($A$) site in perovskites.
Present results and conclusions could be applicable to or serve as a
starting point of investigation for several equivalent problems
where the roles played by chemical variation or thermal
expansion/contraction can be seen as analogous to pressure, as with
spin excitation in rare-earth cobaltites at finite (but low) temperatures.

\bigskip

\textbf{Acknowledgements} This work was primarily supported by the
National Science Council (NSC) of Taiwan under Grant No. NSC
102-2112-M-008-001-MY3 (H.H.) and NSF Awards EAR-1319361, -1019853,
and -0810272 (R.M.W). Calculations were performed at the Minnesota
Supercomputing Institute (MSI) and the National Center for
High-performance Computing (NCHC) of Taiwan.

\newpage

\newpage

\newpage

\newpage

\newpage

\newpage

\newpage

\newpage

\end{document}